\documentclass{sig-alternate}

\usepackage{amsmath}
\usepackage{amssymb}
\usepackage[np,autolanguage]{numprint}
\usepackage{units}
\usepackage{balance}
\usepackage{subfigure}

\usepackage{algorithm} 
\usepackage[noend]{algorithmic} 

\usepackage[pdftex,pdfpagelabels=false]{hyperref} 
\hypersetup{
    unicode=false,          
    pdftoolbar=true,        
    pdfmenubar=true,        
    pdffitwindow=true,      
    pdfstartview={FitV},    
    pdftitle={Declarative Machine Learning -- A Classification of Basic Properties and Types},    
    pdfauthor={}, 
    pdfsubject={},   
    pdfcreator={},   
    pdfproducer={}, 
    pdfkeywords={}, 
    pdfnewwindow=true,      
    bookmarksnumbered=true, 
    bookmarksopen=true,     
    bookmarksopenlevel=2,   
    colorlinks=false,        
    linkcolor=red,        
    citecolor=green,         
    filecolor=black,        
    urlcolor=black          
}

\newtheorem{prop}{Property}

\newenvironment{itemize*}{\begin{itemize}\setlength{\itemsep}{2.5pt}\setlength{\parskip}{0pt}\setlength{\parsep}{0pt}}{\end{itemize}}
\newcommand{\eat}[1]{}
\begin{document}

\title{Declarative Machine Learning --\\ A Classification of Basic Properties and Types}

\numberofauthors{1}
\author{
  \alignauthor Matthias Boehm,~~~Alexandre V.~Evfimievski,~~~Niketan Pansare,~~~Berthold Reinwald\\~\\
  \affaddr{IBM Research -- Almaden;~~San Jose, CA, USA}
}

\maketitle

\pagenumbering{arabic}

\sloppy

\begin{abstract}
Declarative machine learning (ML) aims at the high-level specification of ML tasks or algorithms, and automatic generation of optimized execution plans from these specifications. The fundamental goal is to simplify the usage and/or development of ML algorithms, which is especially important in the context of large-scale computations. However, ML systems at different abstraction levels have emerged over time and accordingly there has been a controversy about the meaning of this general definition of declarative ML. Specification alternatives range from ML algorithms expressed in domain-specific languages (DSLs) with optimization for performance, to ML task (learning problem) specifications with optimization for performance and accuracy. We argue that these different types of declarative ML complement each other as they address different users (data scientists and end users). This paper makes an attempt to create a taxonomy for declarative ML, including a definition of essential basic properties and types of declarative ML. Along the way, we provide insights into implications of these properties. We also use this taxonomy to classify existing systems. Finally, we draw conclusions on defining appropriate benchmarks and specification languages for declarative ML.
\end{abstract}

\section{Introduction}
\label{sec:intro}

Large-scale machine learning (ML) leverages large data collections for advanced analytics in order to find interesting patterns and train robust predictive models. Traditional frameworks and tools like R, Matlab, Weka, SPSS, or SAS provide rich functionality but---except for dedicated packages---struggle to provide scalable analytics. Due to the data-intensive characteristics, increasingly often data-parallel frameworks like MapReduce \cite{DeanG04}, Spark \cite{ZahariaCDDMMFSS12}, or Flink \cite{AlexandrovBEFHHKLLMNPRSSHTW14} are used for cost-effective parallelization on commodity hardware. However, large-scale computation inherently increases the complexity of specifying ML algorithms, especially with regard to efficient and scalable execution.

\textbf{Large-Scale ML Libraries:} Large-scale ML libraries like MLlib (aka SparkML) \cite{MengBYSVLFTAOXX15}, Mahout \cite{mahout}, and MADlib \cite{CohenDDHW09,HellersteinRSWFGNWFLK12} are currently the predominant tools for large-scale ML. These libraries provide algorithms with fixed distributed runtime plans and often expose the underlying physical data representation. Although such libraries are very valuable tools for end-users, it takes substantial effort to write new or customize existing algorithms because it requires knowledge of ML algorithms, their distributed implementation, and the underlying data-parallel framework. Similarly, improvements often require a modification of all individual algorithms to exploit these improvements.

\textbf{Declarative ML:} Declarative ML aims at a high-level specification of ML tasks or algorithms to simplify the usage and/or development of ML algorithms by separating application or algorithm semantics from the underlying data representations and execution plans. Table~\ref{tab:types} categorizes types of declarative ML and delineates them from ML libraries. Overall, the major benefits of declarative ML are: 
\begin{itemize*}
  \item \emph{Simple, Analysis-Centric Specification},
	\item \emph{Physical Data Independence},
	\item \emph{Automatic Execution Plan Generation} (optimization, platform independence, data-size independence),
	\item \emph{Ease of Deployment} (platform independence, adaptivity of ``packaged'' applications), and
	\item \emph{Separation of Concerns} (skill sets of users/devs).
\end{itemize*}
Over time, systems at different levels of abstraction have been proposed by industry and academia. Example systems range from UDF-centric ML extensions of data-parallel frameworks, to domain-specific languages (DSLs) for ML tasks or ML algorithms. This broad spectrum of systems aiming for declarative ML naturally led to a controversy regarding the scope of declarative ML. Not surprisingly---as many ML algorithms are iterative---the discussion centers around the syntax for specifying loops and control flow in general. Various projects adopt an R/Python-like syntax \cite{AbadiABBCCCDDDG16, Lyubimov, GhotingKPRSTTV11, YuSC15, ZhangKR14, ZhangHY09} inheriting the full flexibility of loops, branches, and functions. Others support loops with (1) more restrictive iteration constructs \cite{ CrottyGDKBCZ15,EwenTKM12}, (2) model updates with implicit convergence checks \cite{BorkarBCRPCWR12}, or encapsulating entire algorithm classes as ML tasks \cite{KraskaTDGFJ13,SparksTHFJK15, ZhangKR14}. We argue that the specific language-level syntax is actually \emph{irrelevant} if the ML task or algorithm specification conforms to a set of basic properties required for declarative ML.

\begin{table}[!t]
\centering \vspace{-0.4cm} \small
  \caption{\label{tab:types}Delineation of Types of Declarative ML.}  
  \begin{tabular}{|c|c|}
	  \hline
		\emph{Declarative ML Tasks} & e.g., MLbase \cite{KraskaTDGFJ13,SparksTHFJK15},\\
		(fixed task) & Columbus \cite{ZhangKR14}, DeepDive \cite{ShinWWSZR15}\\
		\hline
		\emph{Declarative ML Algorithms} & e.g., OptiML \cite{SujeethLBRCWAOO11}, SciDB \cite{Brown10,StonebrakerBPR11}\\
		(fixed algorithm) & SystemML \cite{BoehmBERRSTT14,GhotingKPRSTTV11}, SimSQL \cite{CaiVPAHJ13}\\
		\hline
		\hline
		Large-Scale ML Libraries & e.g., MLlib \cite{MengBYSVLFTAOXX15}, Mahout \cite{mahout},\\
		(fixed plan) & MADlib \cite{CohenDDHW09,HellersteinRSWFGNWFLK12}, ORE, Rev R\\
		\hline
  \end{tabular}
	\normalsize	
  \vspace{-0.1cm}
\end{table}

\enlargethispage{\baselineskip}

\textbf{Contributions and Structure:} The primary contribution of this paper is a systematic analysis and classification of declarative machine learning. We first define---in an syntax-independent manner---a set of basic properties in Section~\ref{sec:props}, that any system for declarative ML should satisfy. Subsequently, we describe the types of declarative ML in Section~\ref{sec:types}. Finally, we use this taxonomy to classify existing systems in Section~\ref{sec:classify}, and draw conclusions for defining appropriate benchmarks and languages in Section~\ref{sec:benchmark}.

\section{Basic Properties}
\label{sec:props}

As a foundation for discussing types of declarative ML, we define essential, basic properties in the three categories of data, operations, and result correctness. We also discuss the implications of the individual properties and provide examples of how Apache SystemML---as a representative system for declarative ML---realizes these properties.

\subsection{Physical Data Independence}
\label{sec:data}

The most significant goal of declarative ML is data independence because it decouples the high-level specification of ML tasks or algorithms from the underlying data representations and related runtime plans and operations. 

\begin{prop}\textbf{Independence of Data Structures:} \label{prop:1}
The data types of inputs, intermediate results, and outputs like matrices or scalars are exposed as abstract data types without access to the underlying physical data representations.
\end{prop}

In the context of declarative ML, \emph{independence of data structures} serves two major purposes. First, abstract data types like matrix hide the decision on distributed vs local data representations. Accordingly, specified tasks or algorithms become independent of data size and deployment context (e.g., distributed computation vs streaming, different runtime backends). Second, abstract data types also hide the physical data representation (e.g., dense/sparse matrices, lossless compression), which allow internal improvements of storage and operation efficiency. 

\begin{prop}\textbf{Independence\,of\,Data\,Flow\,Properties:} \label{prop:2}
Data flow properties are not exposed, i.e., the user has, at specification level, no explicit control over properties like partitioning, caching, and blocking configurations.
\end{prop}

The property of \emph{independence of data flow properties} further restricts the notion of abstract data types, disallowing the explicit specification of interesting data flow properties. Examples are (1) caching and checkpointing (e.g., local and distributed caching, with certain storage levels), (2) logical and physical partitioning (e.g., row/column/block partitioning and range/hash partitioning of distributed data sets), (3) blocking configurations (row/column block sizes, fixed or variable logical/physical block sizes), as well as (4) data formats (text or binary cell/block formats). Note, however, that it is valid to allow the specification of ordering because it is both a logical and physical data flow property. 

\textbf{Example SystemML:} SystemML satisfies both properties by exposing only the abstract data types frame, matrix, and scalar without their physical data structures or interesting data flow properties as shown in Figure~\ref{fig:sysml}, as an example of a valid specification. The decisions on physical data flow properties are, however, crucial for performance. Hence, the system automatically injects, for example, caching and partitioning directives via rewrites. Overall, data independence allowed us to evolve and rebase SystemML without changing a single ML algorithm. Examples are extensions such as the support for different sparse representations, compression, and additional backends like Spark or GPUs. In contrast, for example, Mahout Samsara \cite{Lyubimov} does not satisfy the properties of data independence because decisions on dense/sparse and distributed/local matrices (e.g., \texttt{drmFromHDFS}, \texttt{collect}) as well as data flow properties like partitioning (e.g., \texttt{par}) and caching (e.g., \texttt{checkpoint}) are exposed to the user as shown in Figure~\ref{fig:samsara}.

\begin{figure}[!t]
  \centering \vspace{-0.1cm} \small
	\subfigure[SystemML] { \label{fig:sysml}
     \begin{minipage}{3.6cm}
	   \raggedright
	   \texttt{X = read("./X");}\\
     \texttt{y = read("./y");}\\
		 \texttt{p = t(X) \%*\% y;}\\ 
     \texttt{w = matrix(0,ncol(X),1);}\\~\\
     \texttt{while(...) \{}\\
     \texttt{~~~q = t(X)\,\%*\%\,X\,\%*\%\,p;}\\
     \texttt{~~~...}\\
     \texttt{\}}\\
	   \end{minipage}} 
	\subfigure[Mahout Samsara] {\label{fig:samsara}
     \begin{minipage}{4.5cm}
	   \raggedright
	   \texttt{var X = drmFromHDFS("./X")}\\
     \texttt{val y = drmFromHDFS("./y")}\\
     \texttt{var p = (X.t \%*\% y).collect}\\ 
     \texttt{var w = dense(...)}\\
		 \texttt{X = X.par(256).checkpoint()}\\
     \texttt{while(...) \{}\\
     \texttt{~~~q = (X.t\,\%*\%\,X\,\%*\%\,p).collect}\\
     \texttt{~~~...}\\
     \texttt{\}}\\
	   \end{minipage}}
	\normalsize
	\vspace{-0.55cm}
  \caption{\label{fig:example}Examples of ML Algorithm Specifications.}
	\vspace{0.1cm}
\end{figure}

\subsection{Operation Semantics}

The second major goal of declarative ML is to specify ML tasks or algorithms using domain-specific, high-level operations with well-defined semantics to simplify algorithm usage or development, and enable efficient evaluation plans. 

\begin{prop}\hspace{-0.07cm}\textbf{Analysis-Centric\,Operation\,Primitives:} Basic operation primitives, common in the target analytics domain, are supported.\ For ML algorithms, this includes linear algebra and statistical functions, whereas for ML tasks, this includes task-specific primitives and models.
\end{prop}

In order to allow declarative ML with simple specification, there is a need for operation primitives that closely resemble a natural description of ML tasks or algorithms at a conceptual level. For ML algorithms this includes linear algebra, aggregations, and statistical functions but specific domains like deep learning might require additional domain-specific operations like convolution. Similarly, for ML tasks this includes task-specific abstractions, operations, and models. For example, for specifying a task like \texttt{classify}, common classification algorithms, loss functions, and parameters should be supported to describe candidates and the optimization objective. The same applies for the task of general-purpose optimization \texttt{optimize}, where one would expect a way to specify gradient/loss functions, and termination conditions. Note that this property excludes systems that are declarative but unrelated to ML. 

\begin{prop}\textbf{Known Semantics of Operation Primitives:} 
The semantics of operation primitives used to specify ML tasks or algorithms are known to the system in terms of knowledge of operation characteristics and equivalences. 
\end{prop}

We require \emph{operational semantics}, where there exists at least one na{\"\i}ve evaluation plan or straightforward mapping. Knowledge of operation semantics is essential for generating efficient evaluation plans---for example, via rewrites and operator selection---from high-level specifications. In the context of ML, operation semantics also cover characteristics like commutativity and associativity, sparse-safeness (correctness of processing only non-zero cells), value repositioning (e.g., reorg operations like transpose, or order), symmetry properties, as well as an understanding of composite operations (e.g., sum-product for matrix multiplication). This meta information might be built into the system or annotated in case of extensible systems. Overall operation semantics allow to reason about equivalences, alternative execution strategies, and costs of these alternatives.

\begin{prop} \textbf{Implementation-Agnostic Operations:} 
The specification of ML tasks or algorithms is independent of the underlying runtime operations. This property prohibits user-defined execution strategies and  parameterization.
\end{prop}

Specifying \emph{implementation-agnostic operations}---i.e., independent of runtime backends, distributed vs local operations, and execution strategies---is related to the properties for data independence (see Subsection~\ref{sec:data}) but with a focus on operations to ensure the flexibility of alternative or hybrid runtime backends, alternative deployments, and optimizations like rewrites and operator selection. Furthermore, avoiding low level parameterization like degree of parallelism or cache blocking ensures independence of the ML tasks or algorithms from workload characteristics (e.g., data size) and underlying hardware infrastructure. 

\vspace{-0.1cm}
\begin{prop} \textbf{Well-Defined Plan Optimization Objective:} ML tasks or algorithms specify their expected results unambiguously, using a well-defined (potentially multi-criteria) objective for execution plan optimization. 
\vspace{-0.1cm}
\end{prop}

To specify ML tasks or algorithms in an unambiguous manner, the specification must exhibit (implicitly or explicitly) a well-defined plan optimization objective.\ This property differentiates major types of declarative ML. If the specification relies on unambiguous operations, the implicit optimization objective is efficiency (runtime or resource requirements). In case of multi-objective optimizations, we further need to define a primary dimension and constraints for the other dimensions. For example, if we aim to optimize for both efficiency and accuracy, we might want to optimize accuracy in terms of a quality measure (e.g., L2 loss on a holdout dataset) as the primary dimension along with constraints on efficiency (e.g., time budget, number of models).

\textbf{Example SystemML:} SystemML satisfies these properties by minimizing execution time (under memory budget constraints per execution context) of specified ML algorithms, composed of linear algebra and statistical operations with well-defined semantics. The known operation semantics are used to propagate dimension and sparsity information through the entire algorithm, compute memory estimates, apply static (size-independent) and dynamic (size-dependent) rewrites, and eventually decide upon alternative physical operators. Implementation-agnostic operations allow fine-grained optimization decisions per operation. Note that sequences of operations like \texttt{t(X)\,\%*\%\,X\,\%*\%\,p} from Figure~\ref{fig:sysml} or specifications like independent foreach loops (\texttt{parfor}) \cite{BoehmTRSTBV14} are \emph{assertions} on semantics and properties of the algorithm rather than imperative execution strategies.

\subsection{Result Correctness}

The properties of data independence and operation semantics are necessary but not sufficient for declarative ML. We further need to define the notion of result correctness. With regard to practicability for distributed computing, we define operation results as equivalent if they are \emph{essentially the same}, i.e., they are algebraically (logically) equivalent, which ignores round-off errors (e.g., due to partial aggregation or alternative evaluation orders of operations). 

\vspace{-0.1cm}
\begin{prop}\textbf{Implementation-Agnostic\,Results:}~The results of ML tasks or algorithms as well as individual operations are equivalent (essentially the same), independent of type and location of underlying runtime operations. 
\vspace{-0.1cm}
\end{prop}

This property further qualifies \emph{implementation-agnostic operations} (P5). In order to produce correct results, independent of optimization decisions, alternative execution strategies need to produce equivalent results no matter if they are executed locally or as distributed operations. To accomplish that for an operation like \texttt{rand} with fixed seed, both local and distributed operations need to consistently generate seeds for fixed-sized blocks from the initially given input seed. Furthermore, this property prohibits, for example, lossy compression to reduce communication overhead.  

\vspace{-0.1cm}
\begin{prop}\textbf{Deterministic Results:} 
A given ML task or algorithm yields equivalent (essentially the same) results for multiple executions over the same input data and configuration. Randomized tasks or algorithms achieve this using pseudorandom number generators.  
\vspace{-0.1cm}
\end{prop}

Deterministic results of operations and ML tasks or algorithms is an important property, especially with regard to fault tolerance, where the same operation might be executed multiple times. Furthermore, it is also the basis for benchmarking ML systems in a systematic manner.

\textbf{Example SystemML:} In SystemML, the properties of result correctness are satisfied via consistent local and distributed as well as deterministic operations. ML algorithms are composed of these operations, lifting the properties of result correctness to algorithm level too. Furthermore, we bound the round-off errors via numerically stable operations (based on Kahan+) \cite{TianTR12} for descriptive statistics and aggregations. SystemML also provides configuration knobs to disable rewrites and operator selection to force strict computation. However, other than for debugging, we have not seen data scientists or end-users making use of that.

\begin{table*}[!t]
\centering \vspace{-0.15cm} \small
  \caption{\label{tab:systems}Classification of Existing Systems wrt Declarative ML Algorithms (Type 1).}
	{(P1 Indep.\ Data Structures, P2 Indep.\ Data Flow Properties, P3 Analysis-Centric Operations, P4 Known Operations,\\ P5 Impl.-Agnostic Operations, P6 Well-Def.\ Optim.\ Objective, P7 Impl.-Agnostic Results, P8 Deterministic Results)}
  \begin{tabular}{|c|c||c|c|c|c|c|c|c|c||c|c|}
	  \hline
		\textbf{Name} & \textbf{Dist.} & \multicolumn{8}{c||}{\textbf{Basic Properties}} & \textbf{Type} & \textbf{Objective} \\
		 &  & \textbf{P1} & \textbf{P2} & \textbf{P3} & \textbf{P4} & \textbf{P5} & \textbf{P6} & \textbf{P7} & \textbf{P8} & & \\ 
		\hline
		\hline
		RIOT \cite{ZhangHY09} &  & \checkmark  & \checkmark  & \checkmark & \checkmark  & \checkmark  & \checkmark  & \checkmark  & \checkmark & 1 & min runtime \\
		OptiML \cite{SujeethLBRCWAOO11} &  & \checkmark & \checkmark & \checkmark & \checkmark & \checkmark & \checkmark & \checkmark & \checkmark & 1 & min runtime\\
		SystemML \cite{BoehmBERRSTT14,GhotingKPRSTTV11} & \checkmark & \checkmark & \checkmark  & \checkmark  & \checkmark  & \checkmark  & \checkmark  & \checkmark  & \checkmark & 1 & min runtime s.t.\ memory constraints\\
		Mahout Samsara \cite{Lyubimov} & \checkmark &   &   & \checkmark & \checkmark  &  & \checkmark  & \checkmark  & \checkmark & N/A & min runtime\\
		Distributed R \cite{VenkataramanBRAS13} & \checkmark &   &   & \checkmark &  &  & \checkmark  & \checkmark  & \checkmark & N/A & min runtime\\
		Cumulon \cite{HuangB013,HuangJBM015} & \checkmark & \checkmark  & \checkmark & \checkmark  & \checkmark  & \checkmark  &  & \checkmark  & \checkmark & N/A & min costs s.t.\ runtime constraints\\
		DMac \cite{YuSC15} & \checkmark &  & \checkmark  & \checkmark & \checkmark  &  & \checkmark  & \checkmark  & \checkmark & N/A & min runtime s.t.\ memory constraints\\
		TensorFlow \cite{AbadiABBCCCDDDG16} & \checkmark & \checkmark & \checkmark & \checkmark &  & \checkmark & \checkmark & & \checkmark & N/A & min runtime s.t.\ resource constraints\\
		\hline
		SciDB \cite{Brown10,StonebrakerBPR11} & \checkmark & \checkmark & \checkmark & \checkmark & \checkmark & \checkmark &  \checkmark & \checkmark & \checkmark & 1 & min runtime\\
		SimSQL \cite{CaiVPAHJ13} & \checkmark & \checkmark & \checkmark & \checkmark & \checkmark & \checkmark & \checkmark & \checkmark & \checkmark & 1 & min runtime\\
    \hline
		ScalOps \cite{BorkarBCRPCWR12} & \checkmark &  & \checkmark &  & & & \checkmark & \checkmark & \checkmark & N/A & min runtime\\
		Tupleware \cite{CrottyGDKBCZ15} & \checkmark &  & \checkmark & & &  & \checkmark & \checkmark & \checkmark & N/A & min runtime\\
		Emma \cite{AlexandrovKKSTK15} & \checkmark &  & \checkmark & & &  & \checkmark & \checkmark & \checkmark & N/A & min runtime\\
	  \hline
  \end{tabular}
	\normalsize	
  \vspace{-0.15cm}
\end{table*}

\section{Types of Declarative ML}
\label{sec:types}

So far we discussed general properties of declarative machine learning, which apply to all types of declarative ML. We now create a taxonomy of  types of declarative ML, namely \emph{declarative ML algorithms} and \emph{declarative ML tasks}. These types refer to fundamentally different concepts and thus, also differ in their scope of specification.

\subsection{Declarative ML Algorithms (Type 1)}

Declarative ML algorithms allow data scientists to write and customize ML algorithms in a declarative manner. This scope requires fine-grained semantics including control flow and data flow, where the core operation primitives are often based on linear algebra or statistical functions and the common optimization objective is to minimize execution time but other objectives such as resource consumption are possible. The algorithm-centric specification defines precise semantics but leaves substantial freedom regarding data representations and execution plan optimization. This abstraction level allows data scientists to encode algorithms as they are most naturally expressed and thus, to quickly exploit the latest algorithmic advances. End users also benefit from simply calling these algorithms in terms of automatic optimization, adaptivity, and portability. Example system categories are DSL-centric, SQL-centric, and UDF-centric systems.

\subsection{Declarative ML Tasks (Type 2)}

In contrast to declarative ML algorithms, declarative ML tasks allow end users (without ML background), to specify ML tasks like \texttt{classify}, \texttt{factorize}, \texttt{optimize} independent of ML algorithm specifics. This coarse-grained scope includes automatic feature and model selection and allows for the optimization of both model accuracy and runtime. Core operation primitives are task-specific, i.e., depending on the task at hand, alternative candidate algorithms, loss functions (measure for goodness of fit), and hyper parameters are supported. Operation semantics are either built-in or annotated at the level of used algorithms or loss functions. The properties of declarative ML need to apply to the core optimization problem of the given ML task not the entire stack of used ML algorithms, as long as they are annotated with relevant properties that allow reasoning about alternative plans and costs. However, relying on declarative ML algorithms provides additional flexibility. Example system categories for declarative ML tasks are general-purpose optimization, model selection, and feature selection.

\section{Systems Classification}
\label{sec:classify}

Given the taxonomy of basic properties and types of declarative ML, we now classify existing systems. There has been some related work on similar classifications, most notably, Kumar et al.\ defined a notion of a model selection management system \cite{MSMS15}, along with categories of ML systems, but primarily focused on coverage of industrial systems rather than specifics of declarative machine learning. Tables \ref{tab:systems} and \ref{tab:systems2} classify---in the scope of declarative ML algorithms and tasks---existing systems with regard to the defined basic properties and types of declarative machine learning. This classification also indicates distributed vs local operations and the used optimization objective. 

\begin{table*}[!t]
\centering \vspace{-0.15cm} \small
  \caption{\label{tab:systems2}Classification of Existing Systems wrt Declarative ML Tasks (Type 2).}  
	{(P1 Indep.\ Data Structures, P2 Indep.\ Data Flow Properties, P3 Analysis-Centric Operations, P4 Known Operations,\\ P5 Impl.-Agnostic Operations, P6 Well-Def.\ Optim.\ Objective, P7 Impl.-Agnostic Results, P8 Deterministic Results)}
  \begin{tabular}{|c|c||c|c|c|c|c|c|c|c||c|c|}
	  \hline
		\textbf{Name} & \textbf{Dist.} & \multicolumn{8}{c||}{\textbf{Basic Properties}} & \textbf{Type} & \textbf{Objective} \\
		 &  & \textbf{P1} & \textbf{P2} & \textbf{P3} & \textbf{P4} & \textbf{P5} & \textbf{P6} & \textbf{P7} & \textbf{P8} &  &  \\
		\hline
		\hline
		Bismarck \cite{FengKRR12} & &  & \checkmark & \checkmark &  &  & \checkmark & & & N/A & min runtime s.t.\ accuracy constraints \\
		TensorFlow \cite{AbadiABBCCCDDDG16} & \checkmark & \checkmark & \checkmark & \checkmark & \checkmark & \checkmark & \checkmark & \checkmark & \checkmark & 2 & min runtime s.t.\ accuracy constraints\\
		\hline
		MLbase \cite{KraskaTDGFJ13,SparksTHFJK15} & \checkmark & \checkmark  & \checkmark & \checkmark  & \checkmark  & \checkmark  & \checkmark  & \checkmark  & \checkmark & 2 & max accuracy s.t.\ runtime constraints\\
		\hline
		Columbus \cite{ZhangKR14} & & \checkmark  & \checkmark  & \checkmark  & \checkmark & \checkmark  & \checkmark  & \checkmark  & \checkmark & 2 & min runtime s.t.\ accuracy constraints\\
		DeepDive \cite{ShinWWSZR15} & & \checkmark  & \checkmark  & \checkmark  & \checkmark & \checkmark  & \checkmark  & \checkmark  & \checkmark & 2 & max accuracy s.t.\ runtime constraints\\
		\hline
  \end{tabular}
	\normalsize	
  \vspace{-0.15cm}
\end{table*}

\subsection{Declarative ML Algorithms}

\textbf{DSL-Centric Systems:} The class of DSL-centric systems focuses on domain-specific languages (DSLs) for ML to simplify the writing of ML algorithms. Early examples of declarative systems are RIOT \cite{ZhangHY09} and OptiML \cite{SujeethLBRCWAOO11}, which provide R and Scala DSLs, respectively. Both focus on single-node computation only which makes it easier to adhere to basic properties of declarative machine learning. SystemML \cite{BoehmBERRSTT14,GhotingKPRSTTV11} covers both single-node and distributed computation (on MapReduce and Spark) and satisfies all eight properties of declarative ML as described throughout this paper. More recent systems like Cumulon \cite{HuangB013,HuangJBM015}, Mahout Samsara \cite{Lyubimov}, DMac \cite{YuSC15}, and TensorFlow \cite{AbadiABBCCCDDDG16} similarly aim for declarative, large-scale ML but struggle to satisfy all properties of declarative ML. 
Cumulon \cite{HuangB013,HuangJBM015} can be seen as a declarative system. However, in a strict sense, the optimization objective is ill-defined (P6) as the hard runtime constraint cannot be satisfied without knowing the number of iterations until convergence. Cumulon still allows users to explore per-iteration trade-offs of monetary cost and runtime in cloud environments.  
Mahout Samsara \cite{Lyubimov}, Distributed~R \cite{VenkataramanBRAS13}, and DMac \cite{YuSC15} are not declarative because they expose physical data structures and distributed operations to the user (P1, P5). Mahout Samsara and Distributed~R require the user to decide between local and distributed matrices and expose data flow properties like caching and partitioning (P2), while DMac exposes dense/sparse data structures. Additionally, Distributed~R executes arbitrary user-defined R functions---i.e., with unknown operation semantics (P4)---per partition.
TensorFlow \cite{AbadiABBCCCDDDG16} is a compelling system but focuses more on extensibility than declarative specification. Accordingly, operations are handled as black-box kernels, i.e., with unknown operation semantics (P4). Also, deep-learning-centric optimizations like lossy compression for communication work very well for noisy data but do not satisfy the property of implementation-agnostic results (P7) required for general-purpose declarative ML. 

\textbf{SQL-Centric Systems:} The class of SQL-centric systems complements the class of DSL-centric systems as both aim for custom analysis algorithms.\ Common SQL-centric systems are array databases and SQL-like ML (e.g., for Bayesian ML).\ These systems often follow the compelling argument of integrating advanced analytics with traditional query processing to simplify data pre-processing and leverage well-understood abstractions for data and operations.\ At the same time, linear algebra operations are either directly supported or emulated.\ A prime example of array databases is SciDB \cite{Brown10,StonebrakerBPR11} which indeed satisfies all basic properties for declarative ML.\ Furthermore, also SimSQL \cite{CaiVPAHJ13}---as an example of Bayesian ML or more broadly stochastic analysis---satisfies all basic properties of declarative ML.

\textbf{UDF-Centric Systems:} There is also a class of UDF-centric systems that evolved bottom up from existing data-parallel frameworks like MapReduce, Spark, or Flink as well as compiler frameworks to simplify large-scale ML. Examples are ScalOps \cite{BorkarBCRPCWR12} (which compiles Scala UDF workflows to datalog programs), Tupleware \cite{CrottyGDKBCZ15} (which compiles workflows of UDFs in various frontend languages to custom distributed programs of native code via LLVM), and Emma \cite{AlexandrovKKSTK15} (which compiles Scala UDF workflows to Spark and Flink programs). These systems cover a great variety of use cases, but they systematically fail to satisfy several properties of declarative ML. First, data independence (P1) is not satisfied as the UDFs are implemented against custom data structures which makes it hard to efficiently support dense/sparse or compressed datasets. Second, the operation semantics of UDFs are by definition unknown (P4), miss support of analysis-centric operations (P3), and the focus on large-scale computation requires the UDF workflows to realize distributed algorithm implementations (P5).

\subsection{Declarative ML Tasks}

As discussed before, systems for declarative ML tasks mostly target end users (not algorithm developers) and automatically optimize for runtime and accuracy of general-purpose optimization or model and feature selection tasks. 

\textbf{General-Purpose Optimization:} 
Bismarck \cite{FengKRR12} provides in-database general-purpose optimization via incremental gradient decent, where users provide UDFs for initialization, transition, and termination. This abstraction covers many algorithms over existing relational data. Similar to UDF-centric systems, however, it does not satisfy the independence of data structures (P1) and known operation semantics (P4). Furthermore, effective parallelization requires either violations of implementation-agnostic and deterministic results (P7, P8) for the pure UDA approach or modifications of the UDFs (P5) for the shared-memory UDA approach. The latter also does not satisfy P7/P8 due to Hogwild!-style \cite{RechtRWN11} model updates. 
TensorFlow \cite{AbadiABBCCCDDDG16} also provides primitives for general-purpose optimization via different optimization algorithms. Similar to Bismarck, users provide UDFs for loss and gradient computation. However, as TensorFlow provides sufficient data abstraction via so-called placeholders, it satisfies the properties of data independence. Furthermore, the operation semantics of \texttt{inference}, \texttt{loss}, and \texttt{training} are known and UDFs can leverage the existing built-in functions which also provide abstractions for gradient computation. Thus, although TensorFlow did not qualify for \emph{declarative ML algorithms}, it satisfies the properties at the level of \emph{declarative ML tasks}.

\textbf{Model and Feature Selection:}
MLbase \cite{KraskaTDGFJ13} with its TUPAQ \cite{SparksTHFJK15} component for automatic model search allows users to specify candidate model configurations, a quality measure, and runtime constraints (e.g., models considered, number of scans, etc) and returns the best model along with tuned hyperparameters. Disregarding the underlying library of ML algorithms, MLbase can be classified as a system for declarative ML tasks, as the model search problem is well-defined, independent of the underlying data and operations, with annotated algorithm characteristics, and deterministic results (unless the runtime constraint is a runtime budget). 
Columbus \cite{ZhangKR14} allows users to specify feature engineering workflows in R including data preparations and model building with existing R packages. The optimization objective is to minimize runtime, but an error tolerance allows for more aggressive reuse by leveraging runtime-accuracy trade-offs. Columbus also satisfies the properties for declarative ML tasks.
DeepDive \cite{ShinWWSZR15} enables knowledge base construction via statistical inference. In a first step, users specify SQL queries and UDFs for feature extraction to populate a factor graph. In a second step, candidate mappings are specified via SQL queries to express rules for entities and relations. Finally, marginal probabilities are learned via statistical inference over this factor graph. As data and operations are abstracted via SQL and the actual inference algorithms and details are not exposed, DeepDive can be classified as a declarative system for feature selection.

\textbf{Other System Categories:} Finally, there are also declarative systems for more specific ML tasks like time series forecasting (e.g., Fa \cite{DuanB07}, a skip-list approach \cite{GeZ08}, or F2DB \cite{FischerRL12}), which also consider the trade-off between accuracy and runtime (model selection) but are not subject to this classification for general-purpose declarative ML.

\section{Benchmarking ML Systems}
\label{sec:benchmark}

Having discussed the individual types of declarative ML, it is clear that there cannot exist a single benchmark to cover them all. Existing benchmarks for large-scale computation like BigBench \cite{BaruBCDFGJJKNPRRRSSYY14,GhazalRHRPCJ13}, SparkBench \cite{AgrawalBDLLRRSS15,LiTWZS15}, or HiBench \cite{HuangHDXH10} do cover machine learning but often simply refer to reference implementations of large-scale ML libraries. This is fine to evaluate underlying Hadoop or Spark implementations but simply cannot serve as a benchmark for declarative ML. 

\textbf{A Case for Type-Specific Benchmarks:} We argue that industry and academia is best served with benchmarks specific to types of declarative machine learning (see Section~\ref{sec:types}). In order to properly reflect common workload characteristics, the benchmarks should be further tailored to major subcategories of systems. For example, systems for \emph{declarative ML algorithms} cover the major sub-categories of DSL-centric, UDF-centric, and SQL-centric systems, which---despite some overlap of operation primitives---all target different primary usage scenarios. This calls for specific benchmarks that allow fair comparisons and foster system advancements via challenging workloads. Interestingly, exactly that already happened for SQL-centric systems as both, the SciDB \cite{Brown10,StonebrakerBPR11} and SimSQL \cite{CaiVPAHJ13} projects published benchmarks covering the main characteristics of array databases \cite{TaftVSSMS14} and Bayesian ML \cite{CaiGLPVJ14}. Defining simple yet challenging ML benchmarks that also cover common workload characteristics is not easy, so we---as a community---should highly appreciate contributions in this area. 

\textbf{Note on Specification Languages:} The same reasoning as with system-type-specific benchmarks also applies to specification languages. In order to satisfy the property of \emph{analysis-centric operation support} (P4), systems need to support the operations of their primary analysis use case, which motivates tailor-made specification languages. However, as the properties for declarative ML are syntax independent, we could establish a common syntax for a specification language to cover multiple types of declarative ML. 

\section{Conclusion}

To summarize, we introduced a taxonomy of declarative ML in terms of basic properties of data, operations, and result correctness as well as types of systems for declarative ML. The classification of existing systems has shown that this taxonomy is indeed a useful tool for qualifying systems characteristics in a systematic manner. Fundamentally, this makes a case for a syntax-independent classification of declarative ML, which disqualifies the philosophical argument against loops and control flow in general. 
We are at the beginning of an exciting era of declarative ML, with a good understanding of various aspects but also lots of open research challenges. As advanced analytics become ubiquitous and technology environments are changing at an increasing rate, a declarative specification of ML tasks or algorithms becomes increasingly important. Accordingly, we encourage the research community to participate in this discussion on basic properties of declarative ML in order to eventually converge to a common understanding.


\small
\bibliographystyle{abbrv}
\bibliography{tr_dml2016}
\normalsize

\end{document}